**М. Л. КАЛУЖСКИЙ**

Омский государственный технический университет

# ОСОБЕННОСТИ ТРАНСФОРМАЦИИ МАРКЕТИНГА В ЭЛЕКТРОННОЙ КОММЕРЦИИ

**Статья о влиянии электронной коммерции на трансформацию теории и практики маркетинга. Автор рассматривает интернет-маркетинг как самостоятельную форму маркетинга, формирующуюся по общим законам в новых институциональных условиях.**

Широкое распространение электронной коммерции неизбежно сказалось на теории и практике маркетинга, изменив не только их форму, но и содержание. Трансформация маркетинга стала результатом неформальной институционализации новых форм экономических отношений. Для теории маркетинга это особенно актуально, так как именно маркетинг имеет дело с рыночной средой предприятия.

Вместе с тем общая теория интернет-маркетинга пока не сформирована. Даже в западной теории маркетинга описание интернет-маркетинга пока ещё носит фрагментарный характер. Например, в базовом учебнике Ф. Котлера, несмотря на признание революционности изменений в маркетинге под влиянием развития электронной коммерции, его описанию отведена лишь одна глава. Причем методы интернет-маркетинга упоминаются там вскользь как частный случай прямых продаж [1, с. 777—804].

Отечественные авторы часто вообще считают интернет-маркетинг частным приложением отдельных функций маркетинга. Например, А. Э. Калинина отождествляет интернет-маркетинг с деятельностью по размещению и анализу электронных публикаций в сети Интернет [2, с. 61—64]. Тогда как И. А. Крымский и К. В. Павлов отождествляют интернет-маркетинг и Интернет-рекламу [3, с. 164—170], а Ф. Ю. Вирин понимает под интернет-маркетингом исключительно «*построение маркетинговых коммуникаций через Интернет*» [4, с. 12].

Подобный подход к интернет-маркетингу можно выразить словами Ф. Ю. Вирина: «*...интернет-маркетинг не существует сам по себе, он лишь часть общего маркетинга компании. Интернет-маркетинг — это инструмент, который решает часть задач маркетинга компании, и не больше*» [4, с. 12].

С такими утверждениями абсолютно нельзя согласиться. Подобная трактовка имела право на существование, когда Интернет был незначительным коммуникационным каналом в традиционной индустриальной экономике. Но когда и сам рынок, и его участники, и методы продвижения продукции переместились в виртуальное пространство, ситуация кардинально изменилась. Интернет-маркетинг превратился в полноценный маркетинг, осуществляемый виртуальными участниками виртуального пространства.

Мало того, в теории маркетинга никогда не было и не могло быть каких-либо ограничений, отрицающих возможность ведения полноценной маркетинговой деятельности в сети Интернет. Маркетинг всегда трактовался не как банальный набор инструментов продаж, а как «*целенаправленный упорядоченный процесс осознания проблем потребителей и регулирования рыночной деятельности*» [1, с. 21].

Аналогичным образом маркетинг определяет и теория менеджмента: «*Маркетинг — отличительная черта, уникальная функция бизнеса. ... Любая организация, в которой маркетинг или отсутствует, или является случайным, не является бизнесом*» [5, с. 47].

В условиях сетевой экономики и виртуальных форм ведения бизнеса интернет-маркетинг представляет собой ту же маркетинговую деятельность, но только в новых условиях и на новом системного уровне самоорганизации. С одной стороны, интернет-маркетинг выступает как неформальный институт сетевой экономики, включающий в себя традиции, обычаи и правила ведения экономической деятельности в новых условиях. С другой стороны, интернет-маркетинг представляет собой неотъемлемую функцию электронной коммерции, отвечающую за все взаимодействия фирмы и внешней среды.

Поэтому, если ранее действительно интернет-маркетинг ассоциировался в основном с интернет-коммуникациями, то сегодня он постепенно приобретает черты самостоятельной научной дисциплины. Например, видный американский теоретик маркетинга П. Дойль относит интернет-маркетинг к четвёртой стадии эволюции маркетинга (по собственной классификации), определяя его как «*управление отношениями с индивидуальными покупателями*».

Такая трактовка кардинально отличается от предыдущей стадии, которую П. Дойль называет «*маркетинг как управление торговыми марками*» [6, с. 423]. Она отражает институциональные изменения экономических взаимоотношений, происходящие под влиянием электронной коммерции. Не случайно классик маркетинга Ф. Котлер также называет эти изменения «*революционными*» [1, с. 781].

По мнению П. Дойля, до появления электронной коммерции управление торговыми марками подразумевало выделение целевых сегментов потребительского рынка, а затем разработку и продвижение бренда отдельно для каждого сегмента. Такой подход был ориентирован на товар, но не на потребителя [6, с. 425].

С развитием в середине 1990-х гг. электронной коммерции появилась возможность «*установить отношения с индивидуальными покупателями и в точности удовлетворять их потребности посредством кастомизированных товаров и специализированных услуг*» [6, с. 425]. Именно эту возможность П. Дойль считает отличительной чертой интернет-маркетинга.

Однако и трактовка П. Дойля также уязвима для критики. Особенность американского маркетингового подхода всегда заключалась в том, что свойства товара и его бренд считались важнее, чем предпочтения потребителей. Считалось, если товар общедоступен, обладает приемлемой ценой и хорошим качеством, то ему гарантирован широкий сбыт на рынке.

Это специфика экономической модели США, которая ориентирована на внешнеэкономическую экспансию и экстенсивное развитие за счёт массового производства потребительских товаров. В таких условиях чем стандартнее и качественнее товар, тем легче его продать массовому покупателю [7, с. 24—25].

На описанных принципах до сих пор построена американская модель сетевой экономики, включение в которую ориентации на целевые группы потребителей в сети Интернете действительно открывает дополнительные возможности. Однако этого абсолютно недостаточно, чтобы превратить сетевую экономику в локомотив экономического прогресса.

Бурное развитие интернет-маркетинга произошло не вследствие смены акцентов маркетинговой политики под влиянием новых технологических возможностей, связанных с сетью Интернет, как об этом пишет П. Дойль [6, с. 425]. У этого процесса были макроэкономические и институциональные причины, выходящие за рамки теории маркетинга. Причины, связанные со структурными изменениями в мировой экономике и ликвидацией дисбалансов между производственными и трансакционными издержками.

Для объяснения этих причин следует обратиться к тем самым структурным изменениям в мировой экономике. Еще 10—15 лет назад продвижение товара происходило по длинной цепи товародвижения (дистрибьюторы, оптовики, розница и т.д.). Владельцы торговых марок контролировали, используя маркетинговую «стратегию втягивания», каналы сбыта и обращали трансакционные издержки в свою прибыль. Финансовые институты (банки и эмитенты ценных бумаг) извлекали прибыль за счёт кредитования участников рынка на всех этапах цепи товародвижения.

Однако при этом производство было сначала переведено, а затем отдано на откуп азиатским производителям, которые не имели доступа к внешним рынкам. В первую очередь, речь шла о Китае [8, с. 270—282]. С развитием электронной коммерции азиатские (в основном китайские) производители получили возможность прямого доступа к зарубежным потребительским рынкам. Сформировался принципиально новый подход к организации маркетинговой деятельности — т.н. «азиатский подход» [7, с. 26—27].

Новый азиатский подход к организации маркетинга был основан на минимизации трансакционных издержек, когда посредством Интернета товар доходит до конечного потребителя, минуя традиционных посредников. У азиатских производителей не было выхода на зарубежную инфраструктуру сбыта, а торговые цепи не подразумевали что-то большее, чем прямые договоры с покупателями.

Электронная коммерция позволила азиатским производителям дойти с прямыми поставками до самых отдалённых уголков земного шара и защитить свой рынок от иностранного влияния. Это стало причиной феноменального роста электронной коммерции в Китае. По данным компании McKinsey & Company, китайский рынок электронной коммерции ежегодно удваивается. К 2016 численность онлайн-покупателей в КНР достигнет 351 млн человек, проникновение Интернета в городах составит 80 %, а объем рынка составит — $345 млрд [9, с. 1].

Феномен Китая объясняется ещё и тем, что китайская экономика и её субъекты не отягощены бременем поддержания традиционных каналов сбыта, в условиях падения совокупного покупательского спроса всё больше напоминающих «чемодан без ручки». Благодаря развитию электронной торговли общий вектор экономического развития переориентировался в сторону уменьшения трансакционных издержек и снижения роли государственного регулирования в экономике.

Данное утверждение одинаково относится ко всем странам и континентам. Например, эмпирические исследования рынка электронной коммерции в США свидетельствуют, что и там электронная торговля демонстрирует большую конкурентоспособность в сравнении с традиционной розничной торговлей [10, с. 9—10].

Как совершенно справедливо отмечает Р. Коуз: «*...в отсутствие трансакционных издержек не имеет значения правовая система: люди всегда могут договориться, не неся никаких издержек, о приобретении, подразделении и комбинировании прав так, чтобы в результате увеличилась ценность производства*» [11, с. 19]. Соответственно, там, где не имеет значения правовая система, возрастает значение правил, определяемых экономической целесообразностью доминирующих участников рынка. Иначе говоря, маркетинговой политикой таких участников. А ещё точнее — интернет-маркетинговой политикой.

Сегодня описанные процессы повсеместно наблюдаются в электронной коммерции, в которой роль внешнего права успешно заменяют внутренние правила поведения. Например, согласно правилам международной платёжной системы PayPal, продавец обязан возместить покупателю стоимость покупки, даже если продавец представил доказательства обмана со стороны покупателя. Законодательные нормы здесь не имеют никакого значения, так как пользователи PayPal при регистрации принимают на себя обязательство соблюдать правила платёжной системы и выполнять решения её арбитража.

Таким образом, электронная торговля не отменяет трансакционные издержки, но сводит их к минимуму. В этом заключается секрет эффективности и самодостаточности интернет-маркетинга, всего того, что Ф. Котлер назвал «проклятием» для определённых секторов розничной торговли [12, с. 144].

Виртуализация розничных продаж привела к тому, что трансакционные издержки интернет-компаний опустились ниже трансакционных издержек традиционных продавцов. Согласно сформулированной Р. Коузом теории фирмы: «*фирмы должны возникать просто для осуществления действий, которые в противном случае совершались бы через рыночные трансакции (разумеется, если внутрифирменные издержки меньше, чем издержки рыночных трансакций)*» [11, с. 12]. Сегодня эта теория устарела для вертикально интегрированных, обособленных от рынка компаний, продвигающих свою продукцию в сети Интернет, но она более чем актуальна для действующих там виртуальных компаний и облачных технологий.

В результате изменились методы конкурентного поведения. Территориальный монополизм продавцов остался в прошлом. Теперь достаточно набрать наименование нужного товара в поисковике, и покупатель немедленно получает десятки, а то и сотни конкурентных предложений. Соответственно, конкуренция переместилась из сферы качества, цены и рекламы в сферу сервиса, информации и ассортимента.

Сегодня в электронной коммерции наблюдаются бурно развивающиеся процессы институционализации, проявляющиеся через закрепление неформальных норм и правил поведения. Этот процесс идёт на уровне покупателей, предпринимателей и малого бизнеса, поскольку их потенциал спроса в совокупности намного превышает потенциал среднего бизнеса.

Технологии электронной коммерции эволюционируют в сторону удешевления услуг, а также повышения доступности покупок и технологий продаж. Поэтому основные дивиденды от электронной коммерции получают производители и покупатели, а не торговые посредники, банки и инвесторы, как это было ранее.

Американские маркетологи М. Кристофер и Х. Пэк выделяют три направления качественных изменений в организации бизнеса в условиях электронной коммерции [13, с. 139]:

1) смена ориентации маркетинга от функций к процессам;

2) смена ориентации маркетинга от товаров к покупателям;

3) смена ориентации маркетинга от прибыли к эффективности.

Эволюционно электронная коммерция находится сегодня на таком этапе институциональных отношений, где их содержание уже обрело институциональную основу, но пока ещё не обрело институциональную форму. Интернет-маркетинг, как качественно новая (системная) форма организации маркетинга, повторяет в своём развитии путь общей теории маркетинга.

Три концепции традиционной теории маркетинга, сформировавшиеся в 1930—1940 гг., характеризуют этапы этого пути: [7, с. 22]

*1. Распределительная концепция маркетинга* — увязывает эффективность маркетинговых мероприятий с обеспечением оптимальной доступности товара на рынке. В условиях недостаточности конкуренции (как в традиционном маркетинге) или постоянного притока новых покупателей (как в интернет-маркетинге) распределение играет решающую роль в организации продаж.

*2. Институциональная концепция маркетинга* — увязывает эффективность маркетинговых мероприятий с эффективностью взаимодействия всех заинтересованных в сделке сторон. Несмотря на отдельные примеры лидирующих в Интернете компаний, общие правила рыночного поведения, делающих эту концепцию общепринятой, пока не выработаны.

*3. Функциональная концепция маркетинга* — увязывает эффективность маркетинговых мероприятий с реализацией стандартного набора общепринятых маркетинговых функций, стратегий и инструментов. Эта концепция приобретёт актуальность в период окончательного становления интернет-маркетинга, когда все основные его инструменты и методы будут описаны, а рынок электронной коммерции сформирует устоявшуюся структуру.

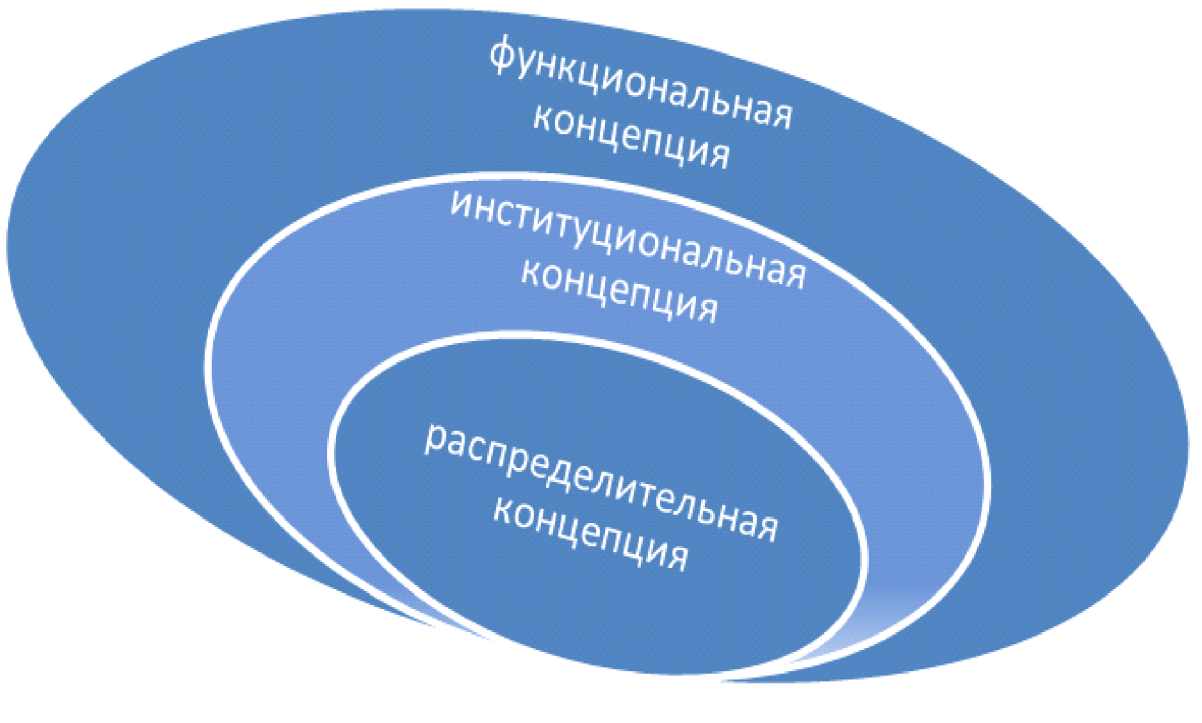

**Рис. 1. Этапы становления маркетинговых концепций**

Все приведённые выше концепции взаимно дополняют друг друга, отражая возрастающую сложность как маркетинговых отношений, так и требований рынка (рис. 1).

В настоящий момент в интернет-маркетинге доминирует распределительная концепция, потенциал дальнейшего развития которой далеко не исчерпан. Практически ежегодно в Интернете появляются новые способы продвижения (социальные сети, блоги и т.д.) и связанные с ними целевые рынки. Опыт их использования ещё только не переходит от количества к качеству. Окончательный переход к доминированию институциональной концепции начнётся тогда, когда теория маркетинга абсорбирует накопленный участниками электронной коммерции опыт.

Поэтому основная задача теории интернет-маркетинга заключается в обобщении, классификации и анализе огромного количества фактического материала, связанного с формирующимися институтами электронной коммерции. Именно здесь сегодня сосредоточено ключевое направление развития не только интернет-маркетинга, но и всей теории маркетинга в целом.

**Библиографический список**

**КАЛУЖСКИЙ Михаил Леонидович,** кандидат философских наук, доцент (Россия), доцент кафедры «Организация и управление наукоёмкими производствами».
Адрес для переписки: frsr@inbox.ru